# Evolving Topics in Federated Learning: Trends, and Emerging Directions for IS Research

*Literature Review*

**Md Raihan Uddin**
LUT University
md.uddin@lut.fi

**Gauri Shankar**
LUT University
gauri.shankar@lut.fi

**Saddam Hossain Mukta**
LUT University
saddam.mukta@lut.fi

**Prabhat Kumar**
LUT University
prabhat.kumar@lut.fi

**Najmul Islam**
LUT University
najmul.islam@lut.fi

## Abstract

Federated learning (FL) is a popular approach that enables organizations to train machine learning models without compromising data privacy and security. As the field of FL continues to grow, it is crucial to have a thorough understanding of the topic, current trends and future research directions for information systems (IS) researchers. Consequently, this paper conducts a comprehensive computational literature review on FL and presents the research landscape. By utilizing advanced data analytics and leveraging the topic modeling approach, we identified and analyzed the most prominent 15 topics and areas that have influenced the research on FL. We also proposed guiding research questions to stimulate further research directions for IS scholars. Our work is valuable for scholars, practitioners, and policymakers since it offers a comprehensive overview of state-of-the-art research on FL.

## Introduction

Federated learning (FL) addresses the critical challenge of data sharing among business organizations, where concerns over privacy and security often create significant barriers. Traditional data sharing methods require organizations to centralize their data, which raises the risk of data breaches, compliance violations, and loss of control over sensitive information. FL in contrast is a distributed and decentralized machine learning (ML) paradigm that enables collaborative model training while preserving data privacy (Li et al. 2020a) and designing secure and privacy preserving information system (IS). FL allows data to remain on local devices, exchanging only model changes to address data privacy, security, and sovereignty concerns (Gupta and Gupta 2023). From a business and organizational perspective, the FL framework plays a significant role in decision-making through local model training, where the organization shares only model parameters to build a distributed collaborative global model; thus, every organizational business data remains protected in its local server (Choudhary et al. 2024). Continuous technological advancement, such as the widespread adoption of the Internet of Things (IoT), has led to an extensive range of applications, (Ficco et al. 2024) from healthcare to smart cities and finance, resulting in a giant expansion in digital data across these industries (Guendouzi et al. 2023). Consequently, there is a growing demand for advanced ML models on FL that prioritize privacy without compromising accuracy (Supriya and Gadekallu 2023). FL has immense potential due to its seamless integration with other technologies and its capacity to facilitate data collection from end devices (Zhang et al. 2021).

Recently, IS scholars have employed their methodological expertise to explore the usage of FL (Choudhary et al. 2024). This exploration has led to the development of innovative Information Technology (IT) artifacts, including constructs, models, methods, instantiations, and design theories. These contributions are pivotal as they not only enrich the IS knowledge base but also provide tools that address complex technological challenges and support the goals of computational design science within IS (Ampel et al. 2024). The computational design science is an interdisciplinary approach that integrates diverse fields to develop novel data representations, computational algorithms, business intelligence and analytics methods, and human computer interaction (HCI) innovations (Rai et al. 2017). The artifacts generated from FL studies are instrumental in this context, as they provide new methodologies and frameworks that can be applied to enhance these areas (Samtani et al. 2022). As a result, there has been a significant increase in the number of articles relevant to IS field, covering a wide range of themes in FL (He et al. 2023; Wang et al. 2024; Zhang et al. 2024). Therefore, evaluating the existing literature to identify new issues and research gaps that may lead to future IS studies is essential. Understanding the dynamic topics and patterns associated with FL can provide precise guidance for IS design researchers.

Indeed, some studies have already been published that use theoretical and systematic literature reviews (SLR) (Choudhary et al. 2024; Guendouzi et al. 2023; Lo et al. 2021; Shanmugarasa et al. 2023) that only focus on narrow topics connected with FL. A few papers were published using bibliometric analysis methods (Farooq et al. 2021; Gupta and Gupta 2023; Velez-Estevez et al. 2022) that had a very limited scope. However, to the our best knowledge, no study has yet conducted a computational literature review (CLR) on FL. This gap motivates us to perform a CLR to map the evolving topics in FL, outline the key trends that have characterized its development, and spotlight the emerging directions poised to define its future. Unlike previous literature reviews, our computational approach leverages data analytics and ML techniques to analyze a vast corpus of academic publications systematically. This enables us to uncover patterns, themes, and revolutions that might not be immediately visible through manual examination, thus offering a comprehensive and subtle understanding of the field's evolution with various topics. Furthermore, with our adopted review approach, we also conducted in-depth exploration in literature to describe the identified topics and propose future research directions for IS researchers. The research questions guiding this investigation are:

**RQ1** What are the patterns and impacts of scholarly publications over recent years, as evidenced by publication output and subject area focus in leading journals?

**RQ2** What are the key trends and evolving topics in federated learning research?

**RQ3** What emerging directions are shaping the future of the field of the IS as revealed through a computational literature review?

By addressing the above questions, we aim to provide a foundational resource for IS researchers entering these domains. Additionally, we intend to assist in the strategic planning of future research agendas and provide valuable insights into the practical uses of FL in different evolving topics. This study aims to not only analyze the current state-of-the-art topics but also pinpoint areas that have not been well examined and possible areas for expansion. The article provides researchers with a framework for navigating and expanding FL boundaries. Specifically, our study spontaneously explores potential research domains such as health care, IoT, cybersecurity, communication, and game theory and contributes to developing innovative information technology artifacts. The next section discuss the theoretical background and then we discuss the research methodology used in this paper. The presentation of the results with their implications and discussion on future IS research is discussed in subsequent sections. Finally, we conclude this paper highlighting the limitations of current approach and a road map for future topic modeling approaches.

## Theoretical Background: FL Literature Review

FL has been the subject of significant scholarly interest in recent years, as evidenced by the rapid growth in published literature every year and its integration with various sectors. The growth of this field, recent years has led to the widespread utilization of SLR (Choudhary et al. 2024; Guendouzi et al. 2023; Lo et al. 2021; Shanmugarasa et al. 2023), bibliometric analysis (Farooq et al. 2021; Gupta and Gupta 2023; Velez-Estevez et al. 2022), and theoretical review (Li et al. 2020a; Supriya and Gadekallu 2023) as review methodologies.

A summary of existing literature reviews on FL is presented in Table 1. Guendouzi et al. (2023) presented a comprehensive review of FL which covered various aspects of FL, including its evolution, categorization, and critical challenges such as communication overhead, data and model heterogeneity, and privacy concerns. The paper also discussed techniques for addressing these challenges and creating efficient tools and frameworks for implementing FL. Another paper by Lo et al. (2021) was investigating FL from a software engineering perspective. In this study provided an overview of the uses of FL in various fields, particularly in sectors that prioritize privacy, such as healthcare and mobile devices. The paper examined different evaluation methodologies and criteria for assessing FL systems, with simulations being the most common technique. A recent paper Choudhary et al. (2024) presented the impact of FL on decision-making within organizations. The study established a connection between FL, game theory, and sustainability and investigated its potential to enhance creativity, problem-solving capabilities, and organizational adaptability based on existing work. It also examined how FL could help address ethical issues such as data privacy and bias and its contribution to decentralized decision-making to achieve sustainability objectives. Another paper Shanmugarasa et al. (2023) was review FL from clients' perspectives. The study highlighted the challenges clients face in FL and provides a detailed analysis of solutions. Innovative approaches such as blockchain for audibility, differential privacy for re-identification prevention, and clustering techniques for personalization are among the solutions analyzed.

Li et al. (2020a) conducted a theoretical review to explore how FL can facilitate privacy-preserving collaborative computation. The study primarily focused on system optimization, security measures, cross-sector applications in healthcare and other industries, and the creation of innovative models that ensure fairness and resilience in the face of data volatility. Another study Supriya and Gadekallu (2023) thoroughly investigated how soft computing techniques improve FL's performance in terms of preserving privacy, enhancing communication efficiency, and managing heterogeneous data in distributed contexts.

Velez-Estevez et al. (2022) used co-word analysis with the aid of SciMAT software, and 23 themes were identified. Finally, they conclude that six themes can potentially develop in FL. The analysis highlights the extensive citation of these themes, suggesting their significant influence on the research community. Another paper Farooq et al. (2021) examined the research profile of FL, incorporating a co-word analysis through VOSviewer to identify evolving research themes. The research focuses on several metrics, such as growth trends, top-cited papers, and productivity measures across authors, institutions, and countries, and identified five main research themes. In another study by Gupta and Gupta (2023) classified the most advanced FL models that utilize Game Theory (GT) techniques. The research outcomes emphasize the significance of GT-based FL models in enhancing profit maximization, authentication, privacy management, trust management, and threat detection.

| Author | Methodology | Research Focus | Research Findings |
|--------|-------------|----------------|-------------------|
| Choudhary et al. (2024) | SLR using the TCCM framework, analyzing 244 articles from 2018-2023 | FL's advancement in theories, research context, characteristics, and methods corresponding to game theories. | Identified data privacy, bias, and accountability in distributed computing through FL. By game theory, FL improved ethical decision-making while maintaining sustainability. |

| Guendouzi et al. (2023) | SLR and 84 published papers from 2019 to 2023 | Comprehensive understanding of FL, evaluation of aggregation methods. | Identified evolution, challenges, aggregation techniques, and tools for efficient FL implementation. |
|---|---|---|---|
| Shanmugarasa et al. (2023) | SLR and 238 published articles from 2017 to 2022 | Focusing on identifying and analyzing the challenges faced by clients and their respective solutions. | Challenges faced by clients including personalization, privacy management, data and device security management, with solution recommendations. |
| Gupta and Gupta (2023) | Bibliometric analysis with SLR papers from 2019 to 2022. | FL models that leverage Game Theory (GT) strategies. | Identified how GT can enhance FL by addressing key issues such as privacy preservation, incentive distribution, and resource management among participating decentralized nodes. |
| Supriya and Gadekallu (2023) | Theoretical review | Use of soft computing techniques to improve FL and application of ML over private data. | Identified soft computing techniques to improve performance of FL for privacy preservation and handling heterogeneous data. |
| Velez-Estevez et al. (2022) | Bibliometric mapping analysis using co-wording over 2,444 articles from 2007 to 2022. | Different themes of FL, interconnection among the themes and performance measures. | Identified 23 FL research themes and identified 6 main areas: telecommunications, privacy and security, computer architecture, data modeling, ML, and applications. |
| Li et al. (2020a) | Theoretical review | Review how FL works and discover the main problems and solutions in FL development. | Identified FL enables privacy preserving computation techniques including addressing security concerns in healthcare and industry and ensuring fairness and robustness. |

Table 1. Summary of recent FL review research comparison

Prior literature reviews were conducted using manual curation and topic exploration without the ability to analyze large volumes of articles. As FL has widely been used in many research domains, a literature review that considers vast amount of articles published in various domains would be useful for identifying future research directions for IS research. Consequently, we adopted the CLR technique. This computational approach incorporates automated tools to collect and handle extensive unstructured data efficiently. This allows for the dynamic synthesis of the literature, continuous monitoring of the progress of topics, detection of emerging trends and patterns, and finding hidden themes. In summary, Table 2 compares our technique and other commonly used methodologies for conducting literature reviews on FL.

| **Studies** | 1 | 2 | 3 | 4 | 5 | 6 | 7 | 8 | 9 | 10 | 11 |
|---|---|---|---|---|---|---|---|---|---|---|---|
| (Choudhary et al. 2024) | ✓ | ✓ | × | × | × | × | ✓ | ✓ | × | ✓ | × |
| (Guendouzi et al. 2023) | ✓ | ✓ | × | × | × | × | ✓ | ✓ | × | ✓ | × |
| (Shanmugarasa et al. 2023) | ✓ | ✓ | × | × | × | × | ✓ | ✓ | × | ✓ | × |
| (Gupta and Gupta 2023) | ✓ | ✓ | × | × | × | ✓ | ✓ | ✓ | × | ✓ | × |

| (Supriya and Gadekallu 2023) | × | ✓ | × | × | × | × | × | ✓ | × | × | × |
| (Velez-Estevez et al. 2022) | ✓ | × | ✓ | ✓ | × | ✓ | × | ✓ | × | ✓ | × |
| (Li et al. 2020a) | × | ✓ | × | × | × | × | × | ✓ | × | × | × |
| This paper | ✓ | ✓ | ✓ | ✓ | ✓ | ✓ | ✓ | ✓ | ✓ | ✓ | ✓ |
| Table 2. Comparison with existing literature review studies ||||||||||||

Note, numeric values indicate as follows: 1 = Quantitative Analysis, 2 = Qualitative Analysis, 3 = Data Collection Automation, 4 = Scalability, 5 = Real-Time Analysis, 6 = Algorithmic Topic Evolution Tracking, 7 = Dynamic Literature Synthesis, 8 = Trends and Pattern Identification, 9 = Unstructured Data Processing, 10 = Data-Driven Insights, 11 = Latent Topic Insight.

# Research methodology

## *Computational Literature Review*

SLRs provide systematic and rigorous approach to synthesize existing scholarly debates and highlight research gaps to guide future research. However, conducting a through SLR is time consuming and labor intensive, and the outcome quickly becomes outdated as new research studies can emerge rapidly within that period (Atkinson 2024). A CLR has come as a solution to this problem. The method has automated the stages of SLR using AI and ML approaches which aims to minimize researcher bias, enhance transparency and reduce repeatability, and decrease the time required to analyze the impact, structure, and content of research within a specific domain (Kunc et al. 2018; Mortenson and Vidgen 2016).

CLR uses ML algorithms for text mining and data analysis to discover the latent relationship within the textual corpus of specific by using topic modeling (Arsenyan and Piepenbrink 2023). Topic modelling is a technique that can identify themes from several structured or unstructured textual content by examining word relationships (Asmussen and Møller 2019). There are various computational topic modelling techniques available for CLR; from there, the most popular topic modelling techniques are Latent Dirichlet Allocation (LDA), Latent Semantic Analysis (LSA), Probabilistic Latent Semantic Analysis (PLSA), Non-Negative Matrix Factorization (NMF) (Abdelrazek et al. 2023). This approach improves understanding of thematic structures with unbiased interpretation, thus streamlining the review process.

## *Review Process*

This CLR study was conducted using a combination of Preferred Reporting Items for Systematic Reviews and Meta-Analyses (PRISMA) (Moher et al. 2010) protocols and the CLR guidelines outlined by (Antons et al. 2023). We make such fusion of the methods because PRISMA retrieves a necessary and sufficient number of papers together and CLR accelerates the process by automating the entire analyses.

### Article Selection and Dataset Creation

We followed the PRISMA protocol for a transparent and rigorous scientific review process that guided data collection, screening, identification, and inclusion decisions and procedures (Principe et al. 2022). We used the Scopus research database to collect literature for our dataset since it is acknowledged as a high-quality data source for systematic reviews (Kunc et al. 2018; Mortenson and Vidgen 2016). The primary search conducted using the keywords such as "Federated learning", and "FL" which yielded a total of 14,864 articles. The inclusion criteria for these articles were: time frame, from 2016 to March 2024; define the subject area; only academic journals and conference papers that are particularly relevant to the domain of FL; publication phase final; and the literature is only English. Finally, we selected 11,014 literature studies for our dataset.

### Data Preprocessing

Data preprocessing before topic modeling is crucial in reducing noise, vocabulary, and computational overhead (Arsenyan and Piepenbrink 2023). First, we converted lowercase letters for all texts data to ensure consistency and prevent the repetition of words strictly based on differences in capitalization. After the initial cleaning process, the texts data underwent *tokenization* (Gupta et al. 2022), which involved segmenting the text into individual words. An essential step in this procedure was to identify and remove stop words, such as "the," "is," "if," and "in," which were minimally relevant to topic modeling. Following, we applied bigrams and trigrams (Arsenyan and Piepenbrink 2023) to comprehend the detailed of particular sentences as they increase the LDA model's ability to consider frequently occurring words as single entities, enhancing the analysis with more profound semantic interpretations. *Lemmatization* was the final step in the preprocessing pipeline, where words were reduced to their base or lemma form (Weismayer 2022). We used the spacy [1] library for lemmatization, and our goal was directed toward the essential parts of speech (nouns, adjectives, verbs, and adverbs) that best reflected the thematic substance of the abstracts. Finally, the text sanitization process was methodically implemented throughout the entire corpus, successfully eliminating unnecessary phrases and resulting in a refined dataset that was optimally prepared for our topical analysis.

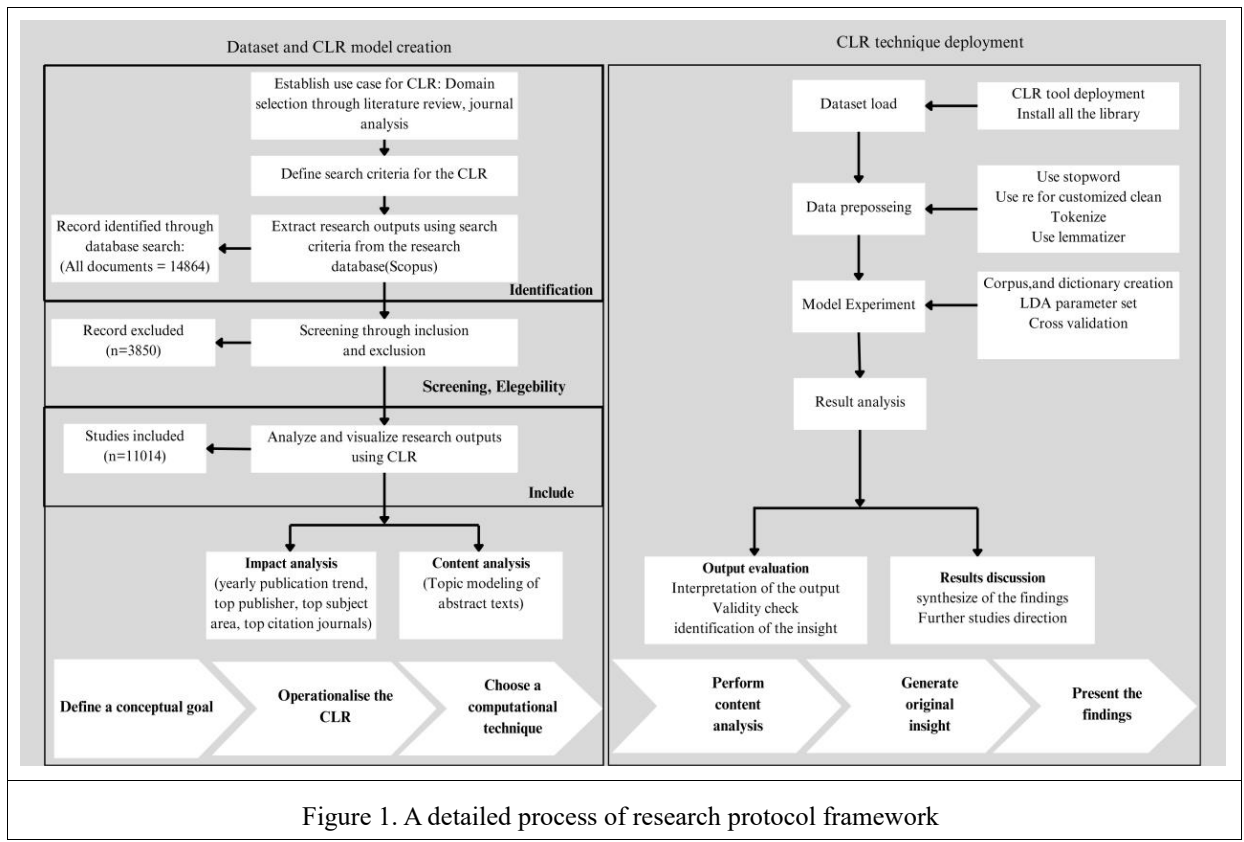

Figure 1. A detailed process of research protocol framework

## Topic Modeling

We use LDA (Maier et al. 2021) as a methodological framework for topic modeling in our CLR. LDA is an unsupervised probabilistic model that generates collections of discrete data, such as text corpora (Chauhan and Shah 2021). We specifically focused on applying LDA with *Gensim* [2] python implementation package. This Python library is optimized for text processing and model implementation for analyzing large datasets. We chose to use LDA due to its widespread popularity as a topic modeling toolkit and its outstanding

---

reliability and precision compared to alternative approaches (Smacchia, Za, et al. 2022). The fundamental idea is that documents are generated from a combination of themes, representing each topic by a word distribution (Vayansky and Kumar 2020). The LDA model is precisely specified by the subsequent generative procedure for every document $w$ in a corpus $D$:

1. Choose $N \sim \text{Poisson}(\xi)$, the number of words in document $w$.

2. Choose $\theta \sim \text{Dir}(\alpha)$, the distribution of topics in document $w$, where $\alpha$ is the parameter of the Dirichlet prior on the per-document topic distributions.

3. For each of the $N$ words $w_n$:
    (a) Choose a topic $z_n \sim \text{Multinomial}(\theta)$.
    (b) Choose a word $w_n$ from $p(w_n|z_n, \beta)$, a multinomial probability conditioned on the topic $z_n$ and $\beta$, the parameter of the Dirichlet prior on the per-topic word distribution.

Figure 1 illustrates the detailed study protocol utilized to investigate the literature using the CLR approach.

## Results

### Impact analysis

One of the main goals of this study is to analyze the evolution trend and scholarly interest in FL by identifying the publication pattern over the years, and the primary research subject areas. Initially, the impact was assessed by analyzing the yearly number of articles during the past nine years and top publication sources. Figure 2 illustrates a significant annual increase in the literature within the field, making a noticeable peak in 2023 and continuing through to March 2024. This trend emphasizes the field's heightened prominence and attractiveness in recent years. By following, Figure 3 presents the distribution of literature across prestigious publications. Notably, "The IEEE Internet of Things Journal" is the prominent source, containing approximately 380 documents. This is closely followed by the "Proceedings of the IEEE Global Communications Conference," which published nearly 300 documents. The sustained growth in scholarly publications in various prestigious journals and conferences reflects the burgeoning interest and investment in this study area. We further assessed the impact by identifying the most relevant area of study. Table 3 shows that Computer Science has the highest number of published papers, totaling 13,137. The areas of engineering, mathematics, decision science, physics and astronomy, and medicine energy are the most prominent in FL.

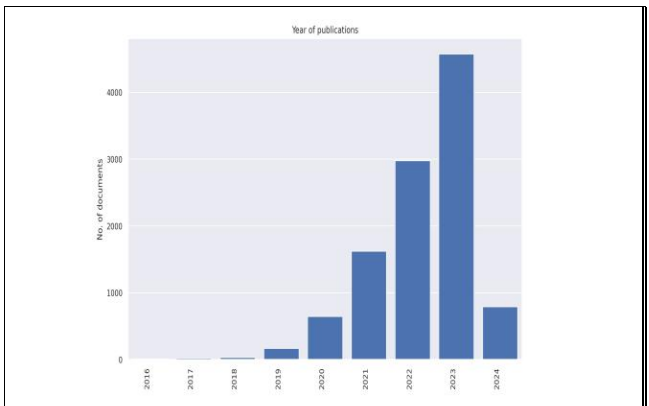

Figure 2. Number of FL publication trend per year

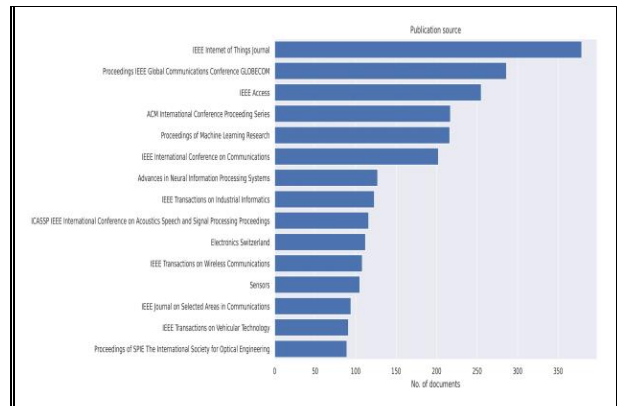

Figure 3. Top publisher for FL research

| Subject area | CS | Eng | Math | DS | PA | Medicine | MS | SS | Energy | BMA | HP |
|---|---|---|---|---|---|---|---|---|---|---|---|
| **Published papers** | 13137 | 6531 | 3448 | 1774 | 917 | 655 | 545 | 539 | 418 | 296 | 175 |
| Table 3. FL published literature work by subject area | | | | | | | | | | | |

Note the abbreviation of CS=Computer Science, Eng=Engineering, Math= Mathematics, DS=Decision science, PA=Physics and astronomy, MS=Material Science, SS =Social science, BMA= Business, management and accounting, HP=Health profession

### Content analysis

Identifying the most suitable number of topics (K) in our model was crucial in our analytical procedure. In order to accomplish this, we utilized a thorough strategy that involved computing coherence scores and perplexity, increased by a meticulous manual analysis of the model's results. A generally suggested guideline rule is to set K equal to the square root of the total number of documents in the corpus (Blei and Lafferty 2007). Using this guideline in our dataset of 11,014 articles indicated an initial estimate of around 105 themes. This heuristic functioned as an initial reference for our investigation, enabling a methodical yet adaptable approach to reveal the thematic organization of our collection of texts. Subsequently, we experimented with several topic ranges, precisely 80, 85, 95, and 120, and different values for document-topic density (alpha) and word-topic-density (eta). Our analysis revealed that the overall coherence, perplexity, and UMass Coherence scores were 0.43, -17.625, and -12.4722, respectively for topic range 90. Further, we analyzed content based on the frequency of topics and topical trends over time.

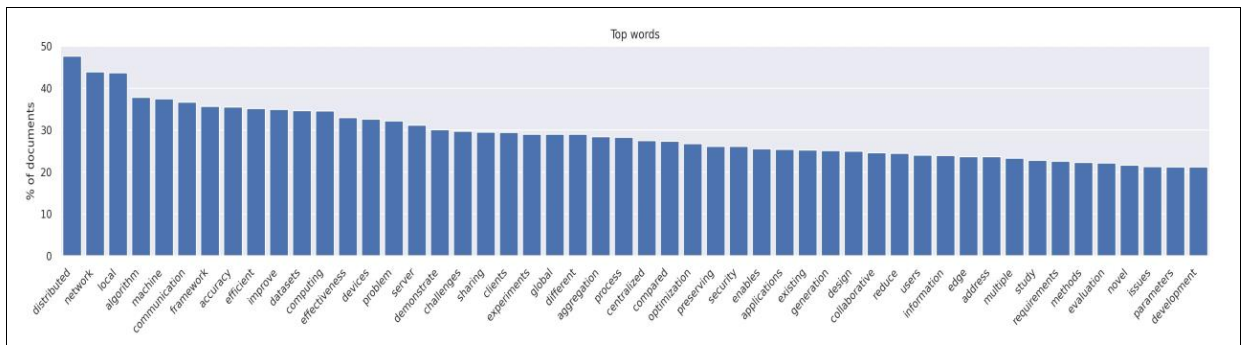

Figure 4. Top words from corpus

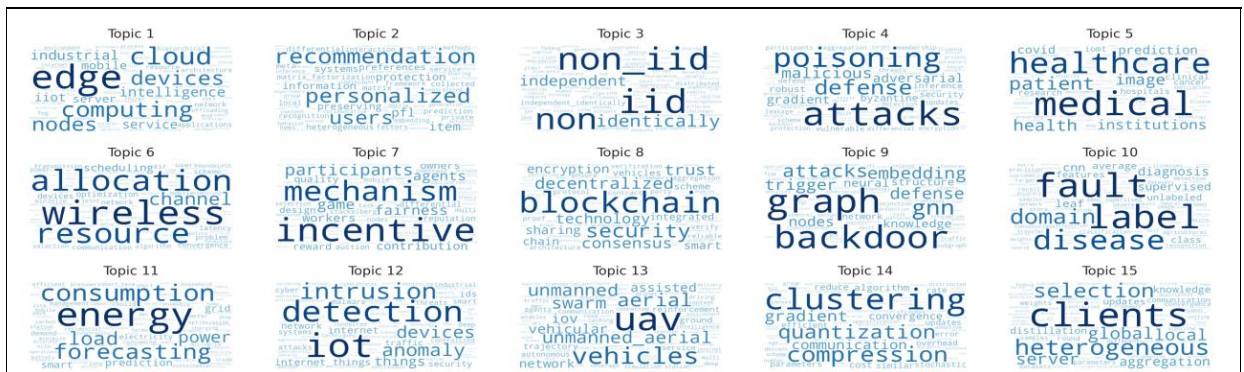

Figure 5. Word cloud of finding topics

**Frequency of topics:** The LDA model produced a series of topics, each defined by a group of prominent words, which indicate the main theme of the topic. To enhance visual clarity and ensure accessibility, Figure 4 presents the frequency distribution of the most significant words across the identified topics. This graphic facilitates the prompt identification of theme areas, emphasizing dominant and prominent topics. Using coherence scores (cv and u-mass) and perplexity scores (Hasan et al. 2021), topic modeling function traverses a range of topic numbers, starting at 15 and ending at 90 with increments of 5. For each number, it constructs a model and computes its coherence score. Figure 6 presents different coherence scores in terms of number of topics. This visual depiction facilitates the identification of the optimal number of topics that achieve maximum coherence. We chose the most suitable model with the highest coherence score (0.46) as optimal based on the computed coherence values. Subsequently, it determines the 15 most excellent topics (see Figure 5) and identifies the most significant words linked with every topic in this model, offering valuable insight into the themes the model represents.

**Trends of topics:** We also conduct an investigation to understand the evolution of topics over the years from 2016 to 2024 on our discovered 15 topics. Figure 7 shows the visual representation of the evolution of the FL based topical patterns among scholarly articles. In recent times (2023-2024), topic related to managing heterogeneous clients shows the highest trends of publishing while edge-cloud computing integration with FL, health care and IoT security & intrusion detection also show dominance in research related to FL. The evolving pattern of the figure also shows that FL was a niche field before the year of 2020, in contrast, after that year FL starts involving in different domains such as ML, game theoretic mechanism, blockchain, graph neural network, etc. and still growing.

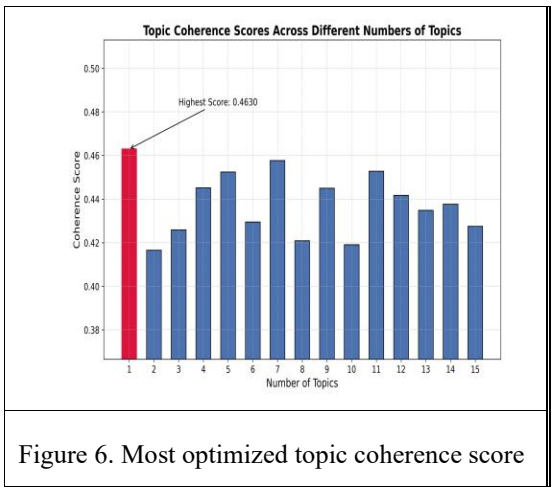

Figure 6. Most optimized topic coherence score

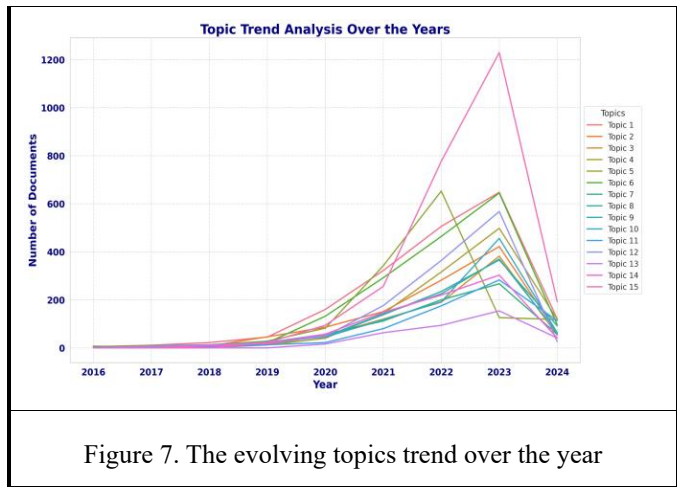

Figure 7. The evolving topics trend over the year

## Discussion

After finishing the impact and content analyses, we further investigated the data by undertaking a comprehensive manual examination of the scholarly literature inside each identified cluster. In order to achieve this objective, we systematically organized and documented essential research papers for each topic, choosing them based on their algorithmically calculated scores of significances. After obtaining the complete text of these articles, we conducted a thorough assessment to determine the thematic content that characterizes each cluster. In the subsequent analysis, we provide concise overviews of each subject, appropriately referencing the most influential sources discovered through our analysis.

### Topic descriptions

Figure 8 shows the identified topics and their broader subdivisions and related keywords.

Topic 1: **Edge-Cloud Computing Integration with FL**: The rising demand for real-time, data-driven decision-making in decentralized environments has brought attention to this topic (Li et al. 2023). FL enriches this integration by allowing edge devices to learn together and create a shared predictive model while localizing all the training data (Lin and Zhang 2022). Integrating edge and cloud computing through FL optimizes computational and data storage capabilities, enabling scalable, efficient, and secure applications across various sectors, including healthcare, autonomous vehicles, and IoT (Cui et al. 2022).

Topic 2: **Personalized Recommendation Systems using FL**: The topic of developing personalized models while protecting user privacy is rising. FL provides decentralized user data-learning recommendation systems without sending sensitive data to a central server (Fan et al. 2023b). This approach aligns with strict data privacy regulations such as the GDPR and improves user trust by minimizing data vulnerability and probable misuse (Sarkar et al. 2023). Practical usage applications are online retail and healthcare platforms while minimizing costs and maintaining high performance.

Topic 3: **Handling Non-IID Data in FL**: It is a widely discussed topic due to its prevalence in real-world applications. It has become increasingly common in many applications like healthcare and IoT, where data privacy and decentralized learning are of utmost importance (Ma et al. 2022; Zhu et al. 2021). As real world scenarios become more complex and data-driven, it is crucial to efficiently manage non-IID data in FL systems to ensure practical model training and performance (Li et al. 2022).

Topic 4: **Adversarial Defense Mechanisms Against Cyberattacks**: The topic is prominent due to increasing complexity of cyber threats, particularly in FL environments where decentralized data becomes a target. Implementing effective defense mechanisms in the application, for instance, critical smart infrastructure ensure that FL systems are robust and secure, protecting them from potential threats that can manipulate, or corrupt the learning process (Nair et al. 2023).

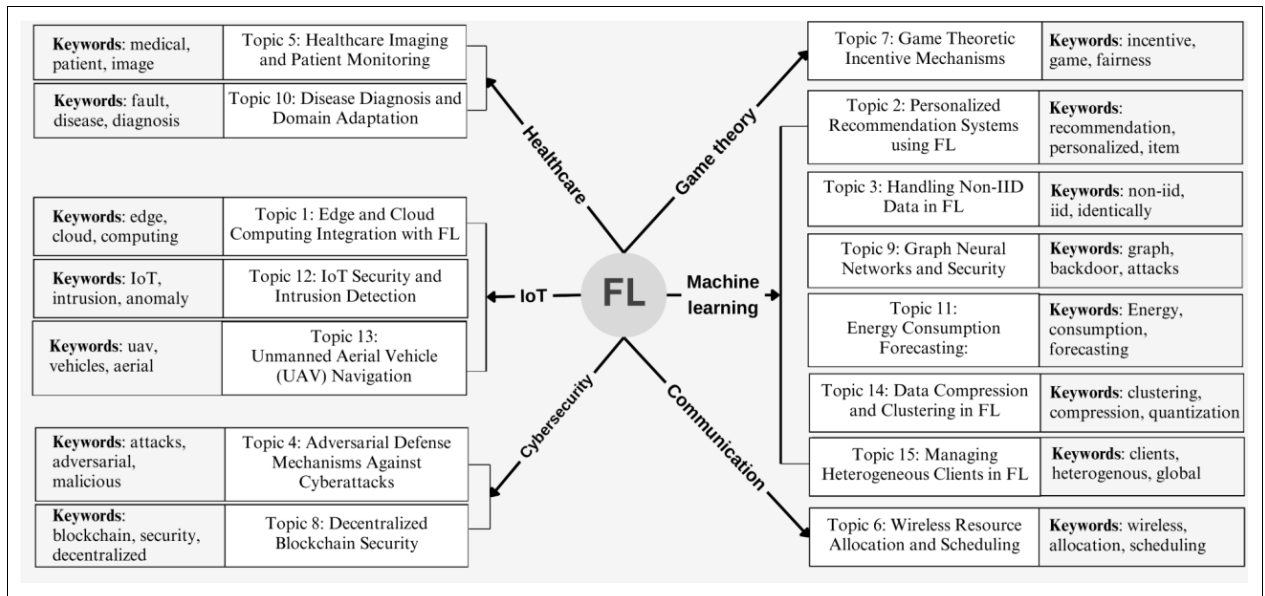

Figure 8. Topic keyword subdivision based on FL

Topic 5: **Healthcare Imaging and Patient Monitoring**: This topic aims to improve medical care by utilizing modern analytics while protecting strict personal data privacy. FL provides a framework in which healthcare data, often sensitive and subject to regulations, can be utilized to enhance diagnostic models and patient monitoring systems without the data ever being transferred from its original place. This technique is crucial in the smart healthcare field, and the scenario of remote and rural medical facilities addresses data privacy and technological difficulties (Nguyen et al. 2022; Ullah et al. 2023).

Topic 6: **Resource Allocation and Scheduling**: The topic has gained prominence because of the increasing need for wireless networks to operate efficiently in various environments for example, smart city traffic management. FL provides a viable solution by facilitating decentralized and collaborative ML while preserving data privacy, which is crucial for wireless networks spanning different geographical and administrative regions (Yang et al. 2019).

Topic 7: **Game Theoretic Incentive Mechanisms**: This topic has highlighted by the need to align the objectives of diverse, autonomous participants within the FL ecosystem (Shi et al. 2020). Game theory provides a structured framework to model the strategic interactions among rational participants, offering insights into the design of incentive mechanisms that can encourage cooperative behavior, optimize resource contribution, and ensure fair distribution of benefits (Huang et al. 2022). Practical usage applications are supply chain management while minimizing fairness and data bias.

Topic 8: **Decentralized Blockchain Security**: The investigation of this topic has become significant due to the rising number of cyber threats that challenge the security of decentralized networks (Fan et al. 2023a). Practical usage applications are supply chain management. With its decentralized and privacy-preserving ML technique, FL promises to improve blockchain security by dynamically enhancing it without centralizing data (Li et al. 2020c; Nguyen et al. 2021).

Topic 9: **Graph Neural Networks and Security**: This topic has gained importance in contemporary research because it can improve ML models on data with a graph structure and ensure the security of distributed learning processes (Mei et al. 2019). FL can offer a decentralized environment to protect data privacy and security in this process, particularly when dealing with the complex structures that Graph Neural Networks (GNNs) handle (Chen et al. 2022). Practical usage applications are social network analysis while supporting flexible network topologies and distributed privacy controls.

Topic 10: **Disease Diagnosis and Domain Adaptation**: The topic has significant interest to the research community due to the critical need for personalized medical diagnostics across diverse patients (Li et al. 2020b). FL presents a novel paradigm that enables the collaborative training of diagnostic models on distributed healthcare data, preserving patient privacy and data confidentiality (Tang et al. 2023).

Topic 11: **Energy Consumption Forecasting**: The topic is gaining significant attention for sustainable energy management practices globally (Saputra et al. 2019). FL offers a decentralized approach to analyzing vast amounts of distributed data from smart grids and IoT devices without compromising privacy, enabling predictive models to forecast energy consumption patterns accurately (Tun et al. 2021).

Topic 12: **IoT Security and Intrusion Detection**: The topic is of burgeoning interest within the cybersecurity and IoT communities (Rahman et al. 2020). As the IoT expands, encompassing many devices, from home appliances to industrial sensors, it becomes a prime target for cyber threats. FL emerges as a novel paradigm to strengthen IoT security by leveraging distributed data sources for anomaly and intrusion detection without compromising data privacy (Campos et al. 2022; Ruzafa-Alcázar et al. 2021).

Topic 13: **Unmanned Aerial Vehicle (UAV) Navigation**: The topic has emerged as a significant area of interest because of the increasing deployment of UAVs in various applications, ranging from agricultural monitoring to emergency response and urban planning (Liu et al. 2020). FL, in this context, offers a promising approach to enhancing UAV navigation systems by leveraging distributed learning to improve decision-making algorithms directly on the UAVs, thus reducing latency and reliance on centralized data processing (Du et al. 2020; Tursunboev et al. 2022).

Topic 14: **Data Compression and Clustering in FL**: Scholars are particularly interested in this topic as it can help reduce the size of updates, thereby alleviating network congestion and enhancing learning speed. Simultaneously, clustering algorithms play a crucial role in grouping similar data points or model updates, which can significantly improve learning efficiency and model accuracy by focusing on more relevant data subsets (Al-Saedi et al. 2021). Practical example is autonomous vehicles to dynamically accommodate heterogeneous data distributions and changing network conditions. The FL framework offers distributed machine-learning capabilities that handle bandwidth and storage constraints in this context (Briggs et al. 2020; Haddadpour et al. 2021).

Topic 15: **Managing Heterogeneous Clients in FL**: The topic has significant attention due to the inherent diversity and disparity among participating devices in FL environment (Nishio and Yonetani 2019). This heterogeneity, manifesting in varied computational capabilities, data distributions across devices, poses substantial challenges in ensuring efficient, fair, and effective learning (Qu et al. 2022). Practical example is cross-border financial services across diverse and heterogeneous client environments where requirement is dynamic client selection, bias-mitigating model aggregation.

### Future research direction for IS research

As described earlier, one of the major IS research areas is designing IT artifact using design science research and computational design science research (Ampel et al. 2024), (Rai et al. 2017), (Samtani et al. 2022). Based on our review, we provide future research directions for IS design research (see Table 4).

| Topic | Theme | Future IS Design Research Directions |
|---|---|---|
| Edge-Cloud Computing Integration with FL | Exploring scalable architectures and improve FL for edge-cloud asynchronous communication. | • How can computational performance be maintained when scaling FL in edge-cloud ecosystems?<br>• How can FL manage asynchronous edge device updates and ensure interoperability? |
| Personalized Recommendation Systems using FL | Improving personalized recommendation systems using FL to address heterogeneity, communication efficiency, and model bias challenges. | • How can FL be optimized in personalized recommendation systems to effectively address communication efficiency, model personalization, and bias issues while minimizing costs and maintaining high performance in diverse market segments? |
| Handling non-IID data in FL | Addressing the challenges of non-IID data in FL to enhance algorithm efficiency, model performance, and data privacy. | • How can FL algorithms be optimized to handle non IID data more effectively while maintaining privacy and performance?<br>• How can FL techniques be designed to efficiently manage non-IID data, improve model performance and security across diverse data distributions? |
| Adversarial Defense Again Cyberattacks | Improving adversarial defense mechanisms in FL to improve robustness against adaptive cyberattacks. | • How can adversarial defense mechanisms in FL be improved to address evolving cyber threats ensuring protection against internal and adaptive cyberattacks while maintaining model performance? |
| Healthcare Imaging and Patient Monitoring | Advancing FL in healthcare imaging and patient monitoring to address data scarcity and privacy issues. | • How can FL optimize healthcare imaging, algorithms, and patient monitoring system personalization and effectiveness while addressing data privacy and technological difficulties in varied medical environments? |
| Resource Allocation and Scheduling | Optimizing wireless resource allocation and scheduling in FL improves model performance and adapts to real world constraints. | • How can FL be adapted to handle wireless resource allocation efficiently and reduce latency to improve the scalability and effectiveness?<br>• How can dynamic wireless resource allocation and scheduling be optimized to enhance |
| | | efficiency and performance in asynchronous multi-model FL? |
| Game Theoretic Incentive Mechanisms | Developing effective incentive mechanisms in FL to address challenges of equity, data heterogeneity, and security in budget constraints. | • How can game theory be utilized to design incentive mechanisms in FL that address client heterogeneity, improve model performance, and ensure fairness, mitigate data bias and profitability? |
| Decentralized Blockchain Security | Improving decentralized blockchain security using FL to address scalability, data integrity, and privacy. | • How can FL be integrated with decentralized blockchain technologies to improve scalability, ensure data integrity, and enhance privacy in IoT and other heterogeneous decentralized devices? |
| Graph Neural Networks and Security | Improving algorithms and encryption techniques strengthen the security and integrity of GNNs through FL. | • How can FL improve GNN security by preserving data privacy and model integrity while supporting flexible network topologies and distributed privacy controls?<br>• How can developing advanced encryption techniques and aggregation methods for FL-integrated GNNs? |
| Disease Diagnosis and Domain Adaptation | Improving domain adaptation for disease diagnosis using FL to manage data heterogeneity and enhance diagnostic robustness. | • How can federated learning be improved to address the challenges of domain adaptation across diverse domains and model accuracy across multiple medical sites while addressing data heterogeneity? |
| Energy Consumption Forecasting | Improving energy consumption forecasting through FL to handle data heterogeneity and improve scalability in dynamic energy markets. | • How can FL be optimized to effectively forecast energy consumption in diverse and dynamic environments, improving prediction accuracy while adapting to the non-IID nature of regional energy data and the evolving demands of renewable energy integration? |
| IoT Security and Intrusion Detection | Developing advanced algorithms and FL-integrated IoT environment to improve security and detect anomaly. | • How can FL for IoT security be designed by addressing non-IID data challenges, improving model performance, and developing efficient aggregation methods through innovative algorithms? |
| Unmanned Aerial Vehicle (UAV) Navigation | Developing FL methods and algorithms for UAV navigation to enhance real time learning efficiency and secure data handling. | • How can FL be designed to meet the challenges of UAV navigation in unpredictable and dynamic settings, ensuring robust, scalable, and secure algorithms for efficient and reliable operation across varied operational scenarios? |
| Data Compression and Clustering in FL | Optimizing data compression and clustering methods through FL in heterogeneous environments. | • How can adaptive data compression and clustering methods be developed and integrated into FL frameworks to dynamically accommodate heterogeneous data distributions and changing network conditions? |
| Managing Heterogeneous Clients in FL | Improving FL through novel client management strategies to address data heterogeneity and | • How can the FL framework integrate advanced client management strategies, such as dynamic client selection, bias-mitigating model |

| | fairness in heterogeneous environments. | aggregation, to improve model performance, fairness, and scalability across diverse and heterogeneous client environments? |
|---|---|---|
| Table 4. Topic, theme, and possible guidance for future IS design research direction | | |

Although most of our presented research questions are design-oriented, we suggest future research on how to link IS theories to the identified research directions. In particular, a prior study (Choudhary et al. 2024) presented IS theories such as game theory, forecasting theory, contract theory, expected utility theory, prospect theory in the FL context that may fit with our identified research directions.

## Conclusion

This CLR has thoroughly examined the constantly evolving field of FL. In doing so, our study contributes to the IS literature in the following ways. First, to the best of our knowledge, our study is one of the very few studies in IS that employed CLR technique. Second, we identified 15 topics and proposed research opportunities for IS research. In particular, our study contributes to the IS literature by proposing several research questions in each of the identified topic areas. These research questions will enrich the research paradigm of design science and computational design science. Third, our review shows the potential of FL in various areas, such as IoT, healthcare, mobile networks, edge computing, cybersecurity, and industrial applications. These indicate that further computational design science research within the IS domain can be conducted in these areas.

Based on our findings, we propose the following practical implication. We found that FL is perhaps most important when designing privacy-preserving IS. Therefore, we advise managers and developers to consider using FL when developing inter-organizational information system to promote data sharing in a privacy preserving manner.

Despite conducting a thorough review, our study suffers from a number of limitations. Due to semantic constraints of LDA, CLR might show static, uncorrelated and non-hierarchical representation of topic distribution and trends. However, to address the issue, word embedding and transformer-based solutions for automatic topic exploration might be promising alternatives. Also, popular meta heuristics approaches such as PSO, ant colony can play an important role to find optimal value without any iterative method.

## Acknowledgement

This work was supported by the Research Council of Finland with CHIST-ERA, grant agreement no - 359790, Di4SPDS-Distributed Intelligence for Enhancing Security and Privacy of Decentralized and Distributed Systems.